\documentclass{article}%
\usepackage{amsmath}
\usepackage{graphicx}%
\usepackage{amsfonts}%
\usepackage{amssymb}
\newtheorem{theorem}{Theorem}
\newtheorem{acknowledgement}[theorem]{Acknowledgement}

\begin{document}

\title{ Tunnelling Effect and Hawking Radiation from a Vaidya Black Hole}
\author{Jun Ren\thanks{thebest2@163.com} \ Jingyi Zhang \ Zheng Zhao$^{\text{ }}$\\Department of Physics, Beijing Normal University, Beijing 100875,\\China}
\maketitle

\begin{abstract}
In this paper, we extend Parikh' work to the non-stationary black
hole. As an example of the non-stationary black hole, we study the
tunnelling effect and Hawking radiation from a Vaidya black hole
whose Bondi mass is identical to its mass parameter. We view Hawking
radiation as a tunnelling process across the event horizon and
calculate the tunnelling probability. We find that the result is
different from Parikh's work because $\frac{dr_{H}}{dv}$ is the
function of Bondi mass $m\left(v\right)$.

PACS number(s): 04.70.Dy

\end{abstract}

\section{Introduction}

Steven Hawking made a striking discovery that basic principles of quantum
field theory lead to the emission of thermal radiation from a classical black
hole\cite{Hawking} in the 1970's. With the emission of Hawking radiation,
black holes could lose energy, shrink, and eventually evaporate completely.
Because black holes radiate thermally, it sets up a disturbing and difficult
problem: information loss paradox. Moreover, when Hawking first proved the
existence of black hole radiation, he described it as tunnelling triggered by
vacuum fluctuations near the horizon. The idea is that when a virtual particle
pair is created just inside the horizon, the positive energy virtual particle
can tunnel out -- no classical escape route exists -- where it materializes as
a real particle. The negative energy particle is absorbed by the black hole,
resulting in a decrease in the mass of the black hole, while the positive
energy particle escapes to infinity, appearing as Hawking radiation. But,
actual derivation of Hawking radiation did not proceed in this way at all.
There were two difficulties to overcome. The first was that in order to do a
tunnelling computation, one needed to find a coordinate system which was
well-behaved at the event horizon. The second was that there did not seem to
be any barrier.

Recently, a method to describe Hawking radiation as a tunnelling process where
a particle moves in dynamic geometry was developed by Kraus and
Wilczek\cite{Kraus} and elaborated upon by Parikh and Wilczek\cite{Parikh1}%
\cite{Parikh2}\cite{Parikh3}. In their method, they take the self-gravitation
into account and it is the tunnelling particle itself that defines the
barrier\cite{Parikh3}. And they also give a leading correction to the emission
rate arising from loss of mass of the black hole corresponding to the energy
carried by the radiated quantum. Following this method, Hemming and
Keski-Vakkuri have investigated the radiation from Ads black
holes\cite{Hemming}, and Medved has studied those from a de Sitter
cosmology\cite{Medved}. In all these investigations, however, all the black
holes are static. In this paper, we will extend the investigation to a
nonstationary black hole -- the Vaidya black hole. There are two crucial
points. First, the Vaidya black hole radiates energy continuously and there is
a problem how to use the condition of the energy conservation. In quantum
mechanics, particle tunnelling a barrier is a instantaneous process, so, the
metric in the coordinates should satisfy Landau' coordinate clock
synchronization condition. Although not all the nonstationary black holes
satisfy Landau' condition, the Vaidya black hole does satisfy it\cite{zhang}.
Consequently, we could fix a certain Bondi time $v^{\prime}$ and at this time
$v^{\prime}$ the Bondi energy of the Vaidya black hole is conserved. Second,
although the components of the metric in the advanced Eddington coordinate
system are not singular, the event horizon and the time like limit surface do
not locate at the same place, which causes difficulty in the calculation of
the emission rate. We have to introduce a new coordinate, $R=r-r_{H}\left(
v\right)  $. In the new coordinate system, the event horizon and the time like
limit surface do locate at the same place. After calculation, the result is
different from Parikh's work because $\dot{r}_{H}$ is the function of $m(v)$.
We emphasize that although the calculation has been employed at the moment
$v^{\prime}$, $v^{\prime}$ is chosen arbitrarily. So, our results reveal the
dynamical change of the non-stationary black hole. Throughout the paper, the
units $G=c=\hbar=1$ are used.

\section{New coordinates}

The metric of a Vaidya black hole can be written as%

\begin{equation}
ds^{2}=-\left[  1-\frac{2m(v)}{r}\right]  dv^{2}+2dvdr+r^{2}(d\theta^{2}%
+\sin^{2}\theta d\varphi^{2}). \label{1}%
\end{equation}
From $g_{00}=0,$ we get, $r=2m(v)$, which is the time like limit surface, not
the event horizon. If we define a new radical coordinate $R=r-r_{H}(v)$, the
line element(\ref{1}) can be rewritten as%

\begin{align}
ds^{2}  &  =-\left[  1-\frac{2m(v)}{r}-2\dot{r}_{H}\right]  dv^{2}%
+2dvdR+r^{2}(d\theta^{2}+\sin^{2}\theta d\varphi^{2})\label{w}\\
&  =\tilde{g}_{00}dv^{2}+2\tilde{g}_{01}dvdR+\tilde{g}_{22}d\theta^{2}%
+\tilde{g}_{33}d\varphi^{2}.\nonumber
\end{align}
The event horizon is\cite{li x}\cite{dai xx}%
\begin{equation}
r_{H}=\frac{2m(v)}{1-2\dot{r}_{H}},
\end{equation}
where $\dot{r}_{H}\equiv\frac{dr_{H}}{dv}$. In the new coordinates, from
$\tilde{g}_{00}=0$ we can obtain the equation of event horizon, $r_{H}%
=\frac{2m(v)}{1-2\dot{r}_{H}}$. So, the event horizon and the time like limit
surface locate at the same place in the new coordinates. The reason why we
must adopt the new coordinates will be explained in Sec.3. Because particle
tunnelling a barrier is a instantaneous process, we will verify that the
metric in new coordinate system satisfy Landau' coordinate clock
synchronization condition as follows.

\bigskip According to Landau' coordinate clock synchronization theory, in a
space-time decomposed in $3+1$, the difference of coordinate times \ \ of two
events taking place simultaneously in different place is%

\begin{equation}
\Delta T=-\int\frac{g_{0i}}{g_{00}}dx^{i},
\end{equation}
where $i=1,2,3$. If the simultaneity of coordinate clocks can be transmitted
from one place to another and have nothing to do with the integration path,
components of the metric should satisfy%

\begin{align}
\frac{\partial}{\partial x^{j}}\left(  \frac{g_{0i}}{g_{00}}\right)   &
=\frac{\partial}{\partial x^{i}}\left(  \frac{g_{0j}}{g_{00}}\right)
,\label{zz}\\
i,j  &  =1,2,3.\nonumber
\end{align}
In this case, $x^{1}=r,x^{2}=\theta,x^{3}=\varphi$, so we can obtain
\begin{equation}
\frac{\partial}{\partial x^{2}}\left(  \frac{g_{01}}{g_{00}}\right)
=\frac{\partial}{\partial x^{1}}\left(  \frac{g_{02}}{g_{00}}\right)
=\frac{\partial}{\partial x^{3}}\left(  \frac{g_{01}}{g_{00}}\right)
=\frac{\partial}{\partial x^{1}}\left(  \frac{g_{03}}{g_{00}}\right)  =0.
\end{equation}

That is to say, we can define coordinate clock synchronization although the
space-time is non-stationary. This feature of the coordinates is very
important for us to discuss the tunnelling process. The area of the event
horizon $A_{H}$, the entropy\cite{li x} of the black hole are given by%

\begin{align}
A_{H}  &  =\int_{r_{H}}\sqrt{-g}d\theta d\varphi=\frac{16\pi m^{2}(v)}{\left(
1-2\dot{r}_{H}\right)  ^{2}},\\
S  &  =\frac{1}{4}A_{H}=\frac{4\pi m^{2}(v)}{\left(  1-2\dot{r}_{H}\right)
^{2}}.\nonumber
\end{align}

To describe across-horizon phenomena, it is necessary to choose coordinates
which are not singular at the horizon; fortunately the line element(\ref{w})
is no singular at the horizon $r_{H}=\frac{2m(v)}{1-2\dot{r}_{H}}$. The radial
null geodesics are given by%

\begin{equation}
\dot{R}\equiv\frac{dR}{dv}=\frac{1}{2}\left[  1-\frac{2m(v)}{r}-2\dot{r}%
_{H}\right]  \label{b}%
\end{equation}
corresponding to outgoing geodesics, where $v$ increases towards the future.
These equations are modified when the particle's self-gravitation is taken
into account. We can consider the particle as a shell of energy. We fix the
total mass (Bondi mass) and allow the hole mass to fluctuate. When the shell
of energy $\omega$ travels on the geodesics, we should replace $m\left(
v^{\prime}\right)  $ with $m\left(  v^{\prime}\right)  -\omega$ in the
geodesic Eq.(\ref{b}) to describe the moving of the shell\cite{Parikh3}.
However, there is a problem that in the nonstationary space-time the energy of
a particle travelling on the geodesics is not conserved because of the
space-time without a time-like killing vecter field. So the energy of shell
$\omega$ will vary when $v$ increases towards the future. The crucial point is
that particle tunnelling a barrier is a instantaneous process. So we can fix a
certain time $v^{\prime}$ and the energy of the black hole is $m\left(
v^{\prime}\right)  $ at the time $v^{\prime}$. When the positive energy
virtual particle with energy $\omega$ just inside the event horizon tunnels
just outside the event horizon where it materializes as a real particle, its
energy is also $\omega$. At the same time the total energy of the black hole
changes to $m\left(  v^{\prime}\right)  -\omega$ and the black hole shrinks a
little. The line element(\ref{w}) should be rewritten as%

\begin{equation}
ds^{2}=-\left\{  1-\frac{2\left[  m\left(  v^{\prime}\right)  -\omega\right]
}{r}-2\dot{r}_{H}\right\}  dv^{2}+2dvdR+r^{2}(d\theta^{2}+\sin^{2}\theta
d\varphi^{2}). \label{c}%
\end{equation}

\section{tunnelling across the horizon}

In our picture, a point particle description is appropriate. Because of the
infinite blue shift near the horizon, the characteristic wavelength of any
wave packet is always arbitrarily small there, so that the geometrical optics
limit becomes an especially reliable approximation. The geometrical limit
allows us to obtain rigorous results directly in the language of particles,
rather than having to use the second-quantized Bogolubov method. When the
black hole varies slowly, in the semiclassical limit, we can apply the WKB
formula. This relates the tunnelling amplitude to the imaginary part of the
particle action at stationary phase. The emission rate, $\Gamma$, is the
square of the tunnelling amplitude\cite{Parikh2}:%

\begin{equation}
\Gamma\thicksim\exp\left(  -2\operatorname{Im}S\right)  .
\end{equation}

The imaginary part of the action for an outgoing positive energy particle
which crosses the horizon outwards from $r_{in}$ to $r_{out}$ can be expressed as%

\begin{equation}
\operatorname{Im}S=\operatorname{Im}\int_{R_{in}}^{R_{out}}p_{R}%
dR=\operatorname{Im}\int_{R_{in}}^{R_{out}}\int_{0}^{p_{R}}dp_{R}^{\prime}dR,
\label{d}%
\end{equation}
where $p_{R}$ is canonical momentum conjugate to $R$. In the coordinates
$\left(  v,R,\theta,\varphi\right)  $, $R_{in}=r\mid_{r_{H}}-r_{H}=0$ is the
initial radius of the black hole, and $R_{out}=\tilde{r}_{H}-r_{H}=-\delta$ is
the final radius of the hole, where $\tilde{r}_{H}=$ $r_{H}\left[  m\left(
v^{\prime}\right)  -\omega\right]  $. We substitute Hamilton' equation
$\dot{R}=\frac{dH}{dp_{R}}\mid_{R}$ into Eq.(\ref{d}), change variable from
momentum to energy, and switch the order of integration to obtain%

\begin{equation}
\operatorname{Im}S=\operatorname{Im}\int_{m}^{(m-\omega)}\int_{R_{in}%
}^{R_{out}}\frac{dR}{\dot{R}}dH=\operatorname{Im}\int_{0}^{\omega}\int
_{R_{in}}^{R_{out}}\frac{2dR}{1-\frac{2\left[  m(v^{\prime})-\omega^{\prime
}\right]  }{r}-2\dot{r}_{H}}\left(  -d\omega^{\prime}\right)  . \label{e}%
\end{equation}
We have used the modified Eq.(\ref{b}) and $H$ is the Bondi energy of the
space-time\cite{Parikh2}. The radiative field do not radiate energy in the
tunnelling process because a certain moment $v^{\prime}$ is fixed to study the
tunnelling process, while the difference of the Bondi energy of the black hole
of two different moments $v_{1}$ and $v_{2}$ is the energy radiated from the
black hole during $v_{2}-v_{1}$. So we could get $dH=d\left[  m\left(
v^{\prime}\right)  -\omega\right]  =-d\omega$ and the minus sign appears. Now
let us explain the problem why we must adopt the new coordinates. If we do not
adopt the new coordinates, the Eq.(\ref{e}) will be%

\begin{equation}
\operatorname{Im}S=\operatorname{Im}\int_{0}^{\omega}\int_{r_{in}}^{r_{out}%
}\frac{2dr}{1-\frac{2\left[  m(v^{\prime})-\omega^{\prime}\right]  }{r}%
}\left(  -d\omega^{\prime}\right)  . \label{v}%
\end{equation}
The integral singularity(the first order pole) is at $r=2\left[  m\left(
v^{\prime}\right)  -\omega^{\prime}\right]  $ where is time like limit
surface, not event horizon. This means that the particle tunnels out of the
time like limit surface, not of the event horizon. However, it is well known
that Hawking radiation comes from the event horizon, not from the time like
limit surface. In Eq.(\ref{e}), $1-\frac{2\left[  m(v^{\prime})-\omega
^{\prime}\right]  }{r}-2\dot{r}_{H}$ equals zero at event horizon. So we adopt
the new coordinate and obtain Eq.(\ref{e}) in order that the integral
singularity is at $r=r_{H}^{\prime}$, where $r_{H}^{\prime}=r_{H}\left(
m\left(  v^{\prime}\right)  -\omega^{\prime}\right)  $. We integrate over $R$
firstly. The integral can be done by deforming the contour, so as to ensure
that positive energy solutions decay in time (that is , into the lower half
$\omega^{\prime}$ plane)\cite{Parikh2}. In this way we obtain%

\begin{equation}
\operatorname{Im}S=\int_{0}^{\omega}\frac{4\pi\lbrack m\left(  v^{\prime
}\right)  -\omega^{\prime}]d\omega^{\prime}}{\left(  1-2\dot{r}_{H}\right)
^{2}}. \label{f}%
\end{equation}

Because $\dot{r}_{H}$ is the function of $\omega^{\prime}$ to the
non-stationary black hole, the integral could not be worked out. However, we
could make the physical meaning clear by the following method. The entropy of
the black hole is $S_{BH}=$ $\frac{4\pi m^{2}\left(  v^{\prime}\right)
}{\left(  1-2\dot{r}_{H}\right)  ^{2}}$\cite{li x} and its derivative of
$m\left(  v^{\prime}\right)  $ is%

\begin{equation}
\frac{dS_{BH}}{dm\left(  v^{\prime}\right)  }=\frac{8\pi m\left(  v^{\prime
}\right)  }{\left(  1-2\dot{r}_{H}\right)  ^{2}}+\frac{16\left[  m\left(
v^{\prime}\right)  \right]  ^{2}}{\left(  1-2\dot{r}_{H}\right)  ^{3}%
}\frac{d\dot{r}_{H}}{dm\left(  v^{\prime}\right)  }=\frac{8\pi m\left(
v^{\prime}\right)  }{\left(  1-2\dot{r}_{H}\right)  ^{2}}-2S_{BH}%
\frac{d}{dm\left(  v^{\prime}\right)  }\ln\left(  1-2\dot{r}_{H}\right)
\label{jj}%
\end{equation}
Doing the integral of $m\left(  \tilde{v}^{\prime}\right)  $, the
Eq.(\ref{jj}) could be written as
\begin{equation}
\int_{m\left(  v^{\prime}\right)  }^{m\left(  v^{\prime}\right)  -\omega
}\frac{8\pi m\left(  \tilde{v}^{\prime}\right)  }{\left(  1-2\dot{r}%
_{H}\right)  ^{2}}dm\left(  \tilde{v}^{\prime}\right)  =\int_{m\left(
v^{\prime}\right)  }^{m\left(  v^{\prime}\right)  -\omega}\frac{dS_{BH}%
}{dm\left(  \tilde{v}^{\prime}\right)  }dm\left(  \tilde{v}^{\prime}\right)
+\int_{m\left(  v^{\prime}\right)  }^{m\left(  v^{\prime}\right)  -\omega
}2S_{BH}\frac{d}{dm\left(  \tilde{v}^{\prime}\right)  }\ln\left(  1-2\dot
{r}_{H}\right)  dm\left(  \tilde{v}^{\prime}\right)  \label{kk}%
\end{equation}
It is obvious that the term of the left-hand side of Eq.(\ref{kk}) is just
$-2\operatorname{Im}S$ in Eq.(\ref{f}) and the first term of the right-hand
side of Eq.(\ref{kk}) is just $\Delta S_{BH}=S_{BH}\left[  m\left(  v^{\prime
}\right)  -\omega\right]  -S_{BH}\left[  m\left(  v^{\prime}\right)  \right]
$. So, the term of the left side of Eq.(\ref{kk}) corresponds to the exponent
of $\Gamma$ and on the right-hand side of Eq.(\ref{kk}) there is a remainder
$\int_{m\left(  v^{\prime}\right)  }^{m\left(  v^{\prime}\right)  -\omega}$
$[-2S_{BH}\frac{d}{dm\left(  v^{\prime}\right)  }\ln\left(  1-2\dot{r}%
_{H}\right)  dm\left(  \tilde{v}^{\prime}\right)  ]$ except $\Delta S_{BH}$,
which is different from Parikh's result.

\section{conclusion and discussion}

The tunnelling rate obtained in this paper is the function of time $v^{\prime
}$ which describes the black hole itself. Moreover, when $m$ is a constant, it
could come back to the result of the static Schwarzschild black hole. So, our
results reflect the dynamical change of the Vaidya black hole. In our opinion,
this result is reasonable and anticipatory. When we study the tunnelling
effect from a black hole, there exists thermal equilibrium between the black
hole and its environment in the case of the static and stationary space-time.
So, Parikh get the result, $\Gamma\thicksim\exp\left(  \Delta S_{BH}\right)  $
\cite{Parikh3}. However, because of the radiation of the black holes the
non-stationary black holes exist more generally. In the Vaidya space-time, the
black hole is non-stationary and it is impossible that the black hole keeps
thermal equilibrium between the black hole and its environment. In addition,
an important difference between stationary and non-stationary black holes is
whether $r_{H}$ is the function of time. Due to non-stationary black holes,
the event horizon $r_{H}$ and the entropy of the black hole are related not
only to the Bondi energy but also to $\dot{r}_{H}$. Consequently, the
variation of $r_{H}$ and the entropy also depends on $\dot{r}_{H}$ when
$m\left(  v^{\prime}\right)  $ changes.

\begin{acknowledgement}
I would like to thank Zhoujian Cao for help. This work is supported by the
National Naturel Science Foundation of China under Grand No.10373003 and the
National Basic Research Program of China (No: 2003CB716300).
\end{acknowledgement}

\end{document}